\documentstyle[12pt]{article}
\topmargin--0cm
\oddsidemargin--1mm
\begin{document}
\renewcommand{\theequation}{\thesection.\arabic{equation}}
\newcommand{\pl}{\partial}
\newcommand{\be}{\begin{equation}}
\newcommand{\ee}{\end{equation}}
\newcommand{\ba}{\begin{eqnarray}}
\newcommand{\ea}{\end{eqnarray}}
\def\R{\relax{\rm I\kern-.18em R}}
\def\1{\relax{\rm 1\kern-.27em I}}
\newcommand{\Z}{Z\!\!\! Z}
\newcommand{\ph}{PS_{ph}}
\def\one{1\hskip-.37em 1}
\def\D{\cal D}
\def\Prod{\prod}
\def\n{\nonumber}
\def\ir{{\rm I}\hskip-.2em{\rm R}}
\def\E{{\rm E}\hskip-.55em{\rm I}}
\def\half{\textstyle{\frac{1}{2}}}
\def\iN{{\rm I}\hskip-.2em{\rm N}}
\def\irtwo{{\rm I}\hskip-.2em{\rm R}^2}
\def\irn{{\rm I}\hskip-.2em{\rm R}^n}
\def\iH{{\rm I}\hskip-.2em{\rm H}}
\def\ra{\rightarrow}
\def\tint{{\textstyle\int}}
\def\d{\partial}
\def\o{\overline}
\def\b{\begin{eqnarray*}}     
\def\e{\end{eqnarray*}}       
\def\bn{\begin{eqnarray}}     
\def\en{\end{eqnarray}}       
\def\<{\langle}
\def\>{\rangle}
\def\dn{d^{n}\!x}
\def\dsx{d^{s}\!x}
\def\dny{d^{n}\!y}
\def\dsy{d^{s}\!y}
\def\{{\lbrace}
\def\}{\rbrace}
\def\D{{\cal D}}
\def\H{{\cal H}}
\bibliographystyle{unsrt}

\begin{center}
{\bf COORDINATE-FREE QUANTIZATION OF FIRST-CLASS CONSTRAINED SYSTEMS}\\
\vskip 0.5cm
{John R. KLAUDER}\\

\vskip 0.5cm
{\em Departments of Physics and Mathematics,
 University of Florida,
Gainesville FL-32611, USA }

\vskip 0.5cm
{Sergei V. SHABANOV} \footnote{on leave from Laboratory of Theoretical
Physics, JINR, Dubna, Russia}\\

\vskip 0.5cm
{\em Department of Theoretical Physics, University of Valencia,
c. Moliner 50, Burjassot (Valencia), E-46100, Spain }
\end{center}

\begin{abstract}
The coordinate-free formulation of canonical quantization,
achieved by a flat-space Brownian motion regularization of phase-space
path integrals, is extended to a special class of closed first-class
constrained systems that is broad enough to include Yang-Mills type
theories with an arbitrary compact gauge group. Central to this extension
are the use of coherent state path integrals and of Lagrange multiplier
integrations that engender projection operators onto the
subspace of gauge invariant states.

\end{abstract}

\section{Introduction}
It is well known that the canonical quantization procedure is consistent
only in Cartesian coordinates \cite{1}. For most physically
relevant systems, it turns out to be possible to find a Cartesian system
of axes and, hence, successfully
apply canonical quantization. Nevertheless, the
Hamiltonian dynamics of a classical system apparently exhibits,
at first sight, a
larger symmetry than the associated canonically quantized system. Indeed,
Hamiltonian equations of motion are covariant under canonical
transformations, while the Heisenberg equations of motion are covariant
 under unitary transformations. Unitary transformations preserve the
spectrum of the canonical quantum operators, while in the classical case
 canonical
transformations do not generally preserve the range of the canonical
variables.

It is worth mentioning in this regard
that the old Bohr-Som\-mer\-feld quantization postulate
\begin{equation}
\oint\limits_{}^{}pdq=2\pi \hbar (n+1/2),\ \ n=1,2,\ldots
\label{1.1}
\end{equation}
is invariant with respect to canonical transformations
\begin{equation}
p\rightarrow {\bar p}(p,q),\ \ q\rightarrow {\bar p}(q,p)
\label{1.2}
\end{equation}
because
\begin{equation}
\oint\limits_{}^{}pdq =\oint\limits_{}^{}{\bar p}d{\bar q}\ .
\label{1.3}
\end{equation}
 As a consequence, since the result is identical in all canonical
coordinate systems, the Bohr-Som\-mer\-feld
quantization is in fact ``coordinate-free".
The characteristic properties of the quantum theory, like the energy
spectrum, will be independent of the choice
of canonical coordinates. In this respect, the old quantum dynamics enjoys
the same symmetry as classical dynamics.

In contrast to the Bohr-Sommerfeld procedure, canonical quantization leads
to a result that is not covariant with respect to the initial
choice of canonical coordinates. For example, for a single degree of freedom,
the coherent-state phase space path integral representation of the
evolution operator
\begin{eqnarray}
&\ &\hskip-1cm\<  p'',q'',t|p',q'\> =\< p'',q''|
e^{-it\H/\hbar}|p',q'\>  \nonumber \\
&= & \int\limits_{}^{} \prod\limits_{\tau =0}^{t}\left(
\frac{dp(\tau )dq(\tau )}{2\pi \hbar}\right)\exp \frac{i}{\hbar}
\int\limits_{0}^{t}d\tau \left[p\dot{q}-h(p,q)\right]\ ,
\label{1.4} \\
\H&=&\tint h(p,q)\,|p,q\>\<p,q|\,dp\,dq/(2\pi\hbar)\ ,
\label{4.2}
\end{eqnarray}
is not covariant with respect to canonical transformations, although
the measure, being the product of local Liouville measures at each moment
of time, is invariant under canonical transformations. The contradiction
follows from the observation that all classical dynamical systems with
positive energy and one
degree of freedom are equivalent to, say, a free particle ($h={\tilde p}^2/2$ after
a suitable canonical transformation). Making such a canonical
transformation in (\ref{1.4}), we seem to arrive at the same conclusion
for quantum systems because the integral (\ref{1.4}) is formally
invariant. Such a conclusion is certainly incorrect.

\subsubsection*{Coordinate-free quantization}

To resolve this paradox, it has been proposed \cite{2} to
interpret the ill-defined path integral (\ref{1.4}) by means of
the regularized expression
\begin{equation}
\int\limits_{}^{}{\cal D}p{\cal D}q\,(\,\cdot\,)
\rightarrow\lim _{\nu \rightarrow
\infty}
{\cal M}_\nu \int\limits_{}^{} {\cal D}p{\cal D}q\,(\,\cdot\,)\, e^{-
\frac{1}{2\nu}\int _{0}^{t}d\tau (\dot{p}^2+ \dot{q}^2)}\ ,
\label{1.5}
\end{equation}
where $\ {\cal M}_\nu $ is a suitable normalization
 and the limit $\nu
\rightarrow \infty$ must be taken {\em after} evaluation of the
path integral. Various factors in (\ref{1.5})
combine to give a Wiener measure
on the two-dimensional
 phase space. In contrast to the integral (\ref{1.4}), the
coherent-state path integral with the regularized measure can be regarded as
a sum over trajectories of a  Brownian particle whose
flat, two-dimensional configuration space
is the original phase space of the system.

The spectrum $E$ of the system can be obtained from
the pole structure of
 the Fourier transform
of the trace of the transition amplitude
\begin{equation} Z_t={\rm
tr}e^{-it\H/\hbar}=\sum\limits_{E}^{}e^{-itE/\hbar}=
\int\limits_{}^{}(dp'dq'/2\pi\hbar)\< p',q',t|p',q'\>\,,
\label{1.6}
\end{equation}
where $\< p',q',t|p',q'\>$ is given by the corresponding path integral.
Under canonical transformations (\ref{1.2}) the
Brownian motion on a flat two-dimensional phase space
remains such a Brownian motion, and if one interprets the stochastic
integral $\int pdq$ in the Stratonovich sense,
 then the spectrum of the system is invariant under canonical
coordinate transformations.

In other words, the coherent-state path integral regularized with the
help of the Wiener measure (\ref{1.5}) provides a ``coordinate-free"
description of quantum theory \cite{2}. Such a regularization
procedure applies to general theories without constraints.

\subsubsection*{Gauge theories}

Hamiltonian path integrals are often used to quantize
 gauge theories \cite{3}. We now have in mind a system of $J$ degrees of
freedom $p=\{p_j\}$, $q=\{q^j\}$, $1\leq j\leq J$. A
main feature of gauge systems is the existence of nonphysical canonical
variables. In the standard formulation,
the formal path integral (\ref{1.4}) is divergent because the
Hamiltonian action for gauge systems is invariant with respect to
transformations
\begin{equation}
q\rightarrow q^\omega ,\ \ p\rightarrow p^\omega
\label{1.7}
\end{equation}
whose parameters $\omega$ depend on the time, that is, there are
 orbits traversed by the gauge transformations (\ref{1.7}) in the phase
space along which the action is constant and
traditionally have an infinite volume. The nonphysical variables can be
associated with these ``gauge" directions in phase space.

To factor out such divergencies of the path integral, one should integrate out
the nonphysical variables and obtain a measure on the physical phase
space
\begin{equation}
[PS]_{ph}=[PS]/{\cal G}\ ;
\label{1.8}
\end{equation}
here ${\cal G}$ consists of all transformations (\ref{1.7}).
Technically, the
procedure amounts to a canonical transformation such that the generators of
(\ref{1.7}) become some elements of a new
set of canonical momenta \cite{3}. This canonical
transformation introduces explicit symplectic coordinates $p^*$ and $q^*$
on the physical phase space (\ref{1.8}). However, it is important to realize
that the canonical coordinates on $[PS]_{ph}$
are themselves defined only up to a
canonical transformation, i.e., the parametrization of the physical phase
space is not unique. As we have argued above, the formal integral in the
Hamiltonian path integral cannot provide a
genuine invariance with respect to
canonical transformations. In the framework of gauge theories, this
invariance implies
gauge invariance because the spectrum of a gauge theory
cannot depend on one or another particular parametrization of the physical
phase space.

Thus, the regularization of the path integral measure with the help of a
Wiener measure and the invariance
under canonical coordinate transformations
it offers should be extended to gauge theories. The aim of this
letter is to address this problem. Hereafter, we use units where $\hbar=1$.

\section{The projection method}
\setcounter{equation}0

\subsubsection*{Special constraint class}

Let $\varphi _a=\varphi _a(p,q)$ be a set of  independent closed first-class
constraints, i.e.
\begin{equation}
\{\varphi _a,\varphi _b\}=f_{abc}\varphi _c\ ,
\label{2.1}
\end{equation}
and for convenience we also suppose that
the Poisson bracket of $\varphi _a$ with the system Hamiltonian
vanishes. The constraints
generate gauge transformations on phase space which in their infinitesimal
form are given by
\begin{eqnarray}
&p &\rightarrow p+\delta p=p+\delta \omega ^a\{p,\varphi_a\}\equiv
p^{\delta \omega } \label{2.2} \\
&q & \rightarrow q+\delta q=q+\delta \omega ^a\{q,\varphi_a\}\equiv
q^{\delta \omega} \  ,\label{2.3}
\end{eqnarray}
for general $\{\omega^a\}$.
From (\ref{2.2}) and (\ref{2.3}) it follows that the infinitesimal
gauge transformations
generated by the constraints are also infinitesimal canonical transformations
\begin{equation}
\{p^{\delta \omega},q^{\delta\omega}\}=\{p,q\}+O(\delta\omega ^2)\ .
\label{2.4}
\end{equation}
A finite gauge transformation can be obtained
by applying the operator $\exp[ -
(\omega ^a ad\, \varphi _a)]$, $ ad\,\varphi _a
=\{\varphi _a,\ \cdot \}$, to phase
space variables.

As noted at the outset, canonical quantization singles out Cartesian
coordinates for special attention. We formulate a special class of
closed first-class constraint systems---which we shall refer to
as constraints of ``Yang-Mills type"---in such a favored set of
coordinates. Specifically, we choose
\begin{equation}
\varphi_a(p,q)=f^j_a(q)p_j\equiv(f_a(q),p)\,,
\label{c.1}
\end{equation}
where $(\,,\,)$ denotes a scalar product in a Euclidean space, and
$f_a(q)$ {\it are linear functions of $q$ chosen so that the
constraints (\ref{c.1}) are of the first class}, i.e. they
satisfy (\ref{2.1}). With this choice, the
gauge transformations (\ref{1.7}) are linear canonical transformations.
It follows for such constraints that
\begin{equation}
p_j\{\varphi_a,q^j\}=\varphi_a(p,q)
\label{c.2}
\end{equation}
holds as an identity, which we shall find useful.
We also assume that there is no operator ordering ambiguity in
the constraints after quantization. This situation is in fact
 entirely realized for a gauge theory based on a compact semi-simple
gauge group $\footnote{The formalism applies also to gauge groups
being the direct product a semi-simple and some number of Abelian
groups.}$.

Such constraints enjoy an additional useful property. If
\begin{equation}
|p,q\>\equiv e^{-iq^jP_j}e^{ip_jQ^j}|0\>\,,
\label{c.3}
\end{equation}
 where $|0\>$ is the ground state
of an harmonic oscillator, i.e, $(Q^j+iP_j)\,|0\>=0$ for all $j$,
denotes the coherent states in the same Cartesian coordinates,
then it follows that
\begin{equation}
e^{-i\Omega^a{\hat\varphi}_a(P,Q)}\,|p,q\>=|p^\Omega,q^\Omega\>\,,
\label{c.4}
\end{equation}
namely the action of any finite gauge transformation is to map one
coherent state into another.  Here $\{{\hat\varphi}_a\}$ denote
the constraint operators that generate the gauge transformations.

\subsubsection*{Coherent state propagator}

The total Hilbert space of a gauge system can always be split
into an orthogonal sum of a subspace formed by gauge invariant states
and a subspace that consists of gauge variant states.
Therefore an averaging over the gauge group automatically leads to
  a projection
operator onto the physical subspace of gauge invariant
states. The physical transition amplitude
is obtained from the unconstrained propagator by averaging
the latter over the gauge group,
\begin{eqnarray} \hskip-.5cm
\< p'',q'',t|p',q'\>^{ph}&\equiv &
\int\limits_{G}^{} \frac{d\mu(\omega)}{Vol\ G} \< p'',q'',t|e^{-i\omega^a
\hat{\varphi}_a}|p',q'\>
 \label{2.6a} \\
&\equiv &
\< p'',q'',t|\hat{P}_G|p',q'\>\\
&=& \int (d^J\!pd^J\!q/(2\pi)^J)
\< p'',q'',t|p,q\>\< p,q|\hat{P}_G |p',q'\>\ ,
\label{2.6b}
\end{eqnarray}
which is a quantum implementation of the classical initial value equation
for first-class constraints.
Here $d\mu(\omega)$ is the invariant measure on the space of gauge group
parameters, and $Vol\ G = \int_G d\mu(\omega)<\infty$ is the gauge group
volume.
In what follows we also adopt a shorthand notation for the
normalized Haar measure
\begin{equation}
\delta \omega \equiv \frac{d\mu(\omega)}{Vol\ G} \ ,
\ \ \ \ \ \int\limits_{G}^{}\delta\omega =1\ .
\label{haar}
\end{equation}
The operator $\hat{P}_G$
is a projection operator onto the gauge invariant subspace.
 Its kernel is determined as the
gauge group average of the unit operator kernel
\begin{equation}
\<p'',q''|p',q'\>^{ph}\equiv\< p'',q''|\hat{P}_G|p',q'\> =
\int\limits_{G}^{} \delta\omega\, \< p'',q''|e^{-i\omega_a
\hat{\varphi}_a}|p',q'\>\ .
\label{2.7}
\end{equation}
For some gauge systems, it can be calculated explicitly as well as
the kernel (\ref{2.6a}) \cite{pr}.

\subsubsection*{The path integral based on the projective method}

Applying the projective formula (\ref{2.6a}) to an infinitesimal
transition amplitude $t\rightarrow \epsilon =t/N$ and making
a convolution of $N$ physical infinitesimal evolution operator
kernels, we arrive at the following representation of the
amplitude (\ref{2.6a})
\begin{eqnarray}
\< p'',q'',t|p',q'\>^{ph} =&\ &
\int \prod\limits_{l=1}^{N-1}(dp^J_ldq^J_l/(2\pi)^J)
\< p'',q'',\epsilon |p_{N-1},q_{N-1}\>^{ph}
 \nonumber\\
&\times &
\< p_{N-1},q_{N-1},\epsilon |p_{N-2},q_{N-2}\>^{ph}
\cdots \< p_1,q_1,\epsilon |p',q'\>^{ph}\  .
\label{3.1}
\end{eqnarray}
In the continuum limit,
where $N\rightarrow \infty,\ \epsilon\rightarrow 0$, while
the product $t=N\epsilon$ is kept fixed,
the convolution (\ref{3.1}) of the kernels (\ref{2.6a})
$(t=\epsilon)$ results
in the coherent state path integral \cite{kl2}
\begin{eqnarray}
\< p'',q'',t|p',q'\>^{ph}&= &{\cal M}
\int {\cal D}C(\omega) {\cal D}p{\cal D}q\, e^{iS_H}\ ,
\label{3.6} \\
S_H
&=&\int\limits_{0}^{t}dt'[\left
(p,\dot{q}) - \omega^a \varphi_a(p,q) - h(p,q)
\right]\ ,
\label{3.7}
\end{eqnarray}
where ${\cal D}C(\omega)= \prod_t \delta\omega(t)$
is a formal  (normalized) measure for the gauge group
average parameters (cf (\ref{2.6a})),
and the symbol $h(p,q)$ is defined in (\ref{4.2}).
Thus, the gauge group averaging
parameters $\omega^a$ become the Lagrange multipliers of the
classical theory in the continuum
limit.

A relation between the path integral (\ref{3.6}) and the projective
formula (\ref{2.6a}) is found in the boundary condition for the
path integral. Recall that the integral (\ref{3.6}) is taken over
phase space trajectories that obey the boundary conditions
\begin{eqnarray}
p(0)&= &p'\ ,\ \ \ \ q(0) = q'\ ; \label{3.9} \\
p(t)&= &p''\ ,\ \ \ \ q(t)=q''\ . \label{3.10}
\end{eqnarray}
It is not hard to  find a gauge transformation such that
\begin{equation} (p^\omega,{\dot q^\omega})-
\omega^a\varphi_a(p^\omega,q^\omega)
= (p,\dot{q})\ .
\label{3.11}
\end{equation}
It is equivalent to solving a linear equation
\begin{equation}
\dot{q}^\omega + \omega^a f_a(q^\omega) = \dot{q}\ .
\label{3.12}
\end{equation}
Having found $q^\omega$ one easily determines $p^\omega$ as its
canonical momenta.

The path integral measure is formally
invariant under canonical transformations
and, hence, the explicit dependence on the Lagrange multipliers of the
action $S_H$ disappears after the canonical transformation constructed
above. The residual coherent state path integral represents a
transition amplitude in the unconstrained Hilbert space. However the integral
$\int {\cal D}C(\omega)$ cannot be factored out because a nontrivial
dependence on the Lagrange multipliers survives at the boundaries.
To maintain the boundary conditions (\ref{3.9}) and (\ref{3.10}),
one can, say, require
\begin{equation}
p^\omega(t) = p''\ ,\ \ \ \ q^\omega(t) = q''\ .
\label{3.13}
\end{equation}
Then it is impossible to satisfy the boundary condition (\ref{3.9})
because equation (\ref{3.12}) admits only one boundary condition,
say, at the final time point. Thus, after the canonical transformation
the path integral
must be taken with boundary conditions that depends on $\omega_a$
\begin{equation}
p^\omega(0) = p'^\Omega\ ,\ \ \ \ q^\omega(0) =q'^\Omega\ ,\ \
\ \ \Omega = \Omega[\omega] \ ,
\label{3.14}
\end{equation}
that is, one gauge group average ``survives" the canonical transformation
that removes the Lagrange multipliers from the action and provides
the equivalence of the path integral (\ref{3.6})  to the projective
representation (\ref{2.6a}).

\section{Gauge fixing and the path integral over physical phase space}
\setcounter{equation}0

In practice, it often turns out to be useful to integrate out the nonphysical
phase-space variables associated with pure gauge degrees of freedom
and work with the path integral over the physical phase space
(\ref{1.8}). For this purpose one usually fixes a gauge \cite{3}
\begin{equation}
\chi_a(q) = 0\ .
\label{g.1}
\end{equation}
By a necessary assumption, each gauge orbit $q^\omega$ must
intersect the gauge condition surface (\ref{g.1}) (at least)
once. Under this assumption a generic configuration $q$ can be
parametrized via lifting it onto the gauge condition surface
along a gauge orbit passing through $q$
\begin{equation}
q = q_\chi^\theta(q^*)\ ,
\label{g.2}
\end{equation}
where $\theta_a$ parametrizes the lift along a gauge orbit, and
points $q=q_\chi(q^*)$ form the surface (\ref{g.1}), i.e., $q^*$
parametrizes the surface (\ref{g.1}).

In the curvilinear coordinates (\ref{g.2}) associated with
the chosen gauge condition, the constraints are linear combinations
of canonical momenta for $\theta_a$, and the Poisson bracket of
the canonical variables $p^*$ and $q^*$ with the constraints vanishes,
that is, $p^*$ and $q^*$ are gauge invariant according to (\ref{2.2})
and (\ref{2.3}).
The $\theta$-dependence
of the action can be absorbed by a shift of the Lagrange
multipliers $\omega^a$
on a suitable linear combination of the velocities
$\dot{\theta}_a$ because the canonical one-form assumes the form
\begin{equation}
p\dot{q} +\omega^a\varphi_a = p^*\dot{q}^* + p_\theta^a\dot{\theta}_a
+\omega^a\varphi_a
\label{g.3}
\end{equation}
and the Hamiltonian is gauge invariant (the $\theta_a$'s are
cyclic variables).

The integral over $\theta_a$ yields the gauge group volume that
cancels the one sitting in the measure ${\cal D}C(\omega)$. Finally,
the integrals over $\omega^a$ and $p_\theta^a$ can also be done, and
one ends up with the integral over physical phase space spanned
by local symplectic coordinates $p^*, q^*$.

This result is usually achieved by a formal restriction of the
path integral measure support in (\ref{3.6}) to a subspace of
the constraint surface $\varphi_a(p,q) =0$ selected by the gauge
(supplementary) condition (\ref{g.1}) \cite{3}:
\begin{equation}
\D p\D q\D C(\omega)e^{-i\tint dt \omega^a\varphi_a}
\rightarrow \D p\D q \prod_t \left(\Delta_{FP}\prod_a
\delta(\chi_a)\delta(\varphi_a)\right)\;,
\label{f.1}
\end{equation}
where $\Delta_{FP} = \det \{\varphi_a,\chi_b\}$ is the Faddeev-Popov
determinant. After the canonical transformation associated with (\ref{g.2})
the Faddeev-Popov measure assumes the form \cite{3}
\begin{equation}
\D p^*\D q^*\D p^\theta\D \theta \prod_t \delta(p^\theta)\delta(\theta)\;,
\label{f.3}
\end{equation}
and the integration over the nonphysical variables $p^\theta$ and
$\theta$ becomes trivial.

Two important observations are in order.  First, the procedure
(\ref{f.1}) corresponds to a canonical quantization {\it after}
the elimination of all nonphysical degrees of freedom (the so called
reduced phase-space quantization). As shown above, the physical
variables are associated with curvilinear coordinates, while
canonical quantization is consistent only in Cartesian coordinates.
As a result canonical quantization and the elimination of
nonphysical degrees of freedom generally do {\em not} commute \cite{christ}.
In other words, the procedure (\ref{f.1}) is not, in general, equivalent to
the Dirac quantization scheme \cite{dir} where nonphysical degrees
of freedom are removed after quantization.

Second, the geometry and topology of gauge orbits may happen to be
such that there exists no unique gauge condition \cite{grib}, meaning
that for any given $\chi_a$ the system
\begin{equation}
\chi_a(q) = \chi_a(q^{\omega_s}) =0
\label{f.2}
\end{equation}
always admits nontrivial solutions with respect to $\omega_s^a$. From
the geometrical point of view, the latter implies that the gauge orbit
$q^\omega$ intersects the gauge fixing surface more than once, namely, at
points $q^{\omega_s}$. Discrete gauge transformations associated with
the gauge variables $\omega_s^a$ do not reduce the number of physical
degrees of freedom, but they do reduce the ``volume" of the physical
configuration and phase spaces. Therefore the formal measure
$\D p^*\D q^*$ can no longer be Euclidean and the corresponding path
integral should be modified. If the residual discrete gauge transformations
are explicitly known, then in such cases it appears to be possible to find
a modified path integral formalism that is equivalent to the
Dirac method \cite{sha2}.

Finally we remark that
the Liouville measure ${\cal D}p^*{\cal D}q^*= \prod_tdp^*(t)dq^*(t)$
is invariant with respect to canonical transformations. This freedom
in the path integral over physical phase space can be interpreted
as  gauge invariance. Indeed, another choice of a gauge condition
(\ref{g.1}) would induce another parametrization of the physical phase
space that is equivalent to the former via a canonical transformation.
On the other hand, we have argued in Section 1 that the formal invariance
of the Liouville measure in the path integral is not sufficient to ensure
the invariance of the quantum theory
with respect to canonical transformations.
In the framework of gauge systems, it implies that, to achieve  gauge
invariance of the path integral over physical phase space, the measure
 should be regularized {\em before} integrating out pure gauge degrees of
freedom with the help of a canonical transformation associated
with a chosen parametrization of the physical phase space by local
symplectic coordinates.

In the next section we propose a generalization of the path integral
measure regularization with a Wiener measure to gauge theories.

\section{The Wiener measure for gauge theories}
\setcounter{equation}0

The Wiener measure regularized phase space path integral for a
general phase function $G(p,q)$ is  given by
\bn
&&\hskip-.3cm\lim_{\nu\ra\infty}{\cal M_\nu}
 \int\exp\{i\tint_0^T[p_j{\dot q}^j+{\dot G}(p,q)-h(p,q)]\,dt\}\n\\
&&\hskip1.5cm\times\exp\{-(1/2\nu)\tint_0^T[{\dot p}^2
+{\dot q}^2]\,dt\}\,\D p\,\D q\n\\
 &&\hskip.3cm=\lim_{\nu\ra\infty}(2\pi)^J e^{J\nu T/2}
 \int\exp\{i\tint_0^T[p_jdq^j+dG(p,q)-h(p,q)dt]\}\,d\mu^\nu_W(p,q)\n\\
&&\hskip.3cm=\<p'',q''|e^{-i\H T}|p',q'\>\;\ ,
\label{4.1}
\en
where the last relation involves a coherent state matrix element.
In this expression we note that $\tint p_j\,dq^j$ is a
{\it stochastic integral}, and as such we need to give it a definition.
As it stands both the It\^o (nonanticipating) rule and the Stratonovich
(midpoint) rule of definition for stochastic integrals yield the same
result (since $dp_j(t)dq^k(t)=0$ is a valid It\^o rule in these
coordinates). Under any change of canonical coordinates,
we consistently will interpret this stochastic integral
in the Stratonovich sense because it will then obey the ordinary
rules of calculus.

Why does the representation of the propagator as well as the Hamiltonian
operator involve coherent states
\bn |p,q\>\equiv e^{-iG(p,q)}e^{-iq^jP_j}
e^{ip_jQ^j}|0\>\;,\hskip1cm(Q^j+iP_j)|0\>=0\;?
\label{4.3}
\en
One simple argument is as follows. The Wiener measure
is on a flat {\it phase space}, and is pinned at both ends thus resulting
in the boundary conditions $p(T),q(T)=p'',q''$ and $p(0),q(0)=p',q'$.
Holding this many end points fixed is incompatible with a Schr\"odinger
representation, which holds just $q(T)$ and $q(0)$ fixed, or with a
momentum space representation, which holds just $p(T)$ and $p(0)$ fixed.
It turns out, as a consequence of the Wiener measure regularization,
that the propagator is {\it forced} to be in a coherent state
representation. We also emphasize the covariance of this expression
under canonical coordinate transformations. In particular,
if ${\o p}d{\o q}=pdq+dF({\o q},q)$ characterizes a canonical
transformation from the variables $p,q$ to ${\o p},{\o q}$,
then with the Stratonovich rule the path integral becomes
\bn
&&\<{\o p}'',{\o q}''|e^{-i\H T}|{\o p}',{\o q}'\>\n\\
&&\hskip.3cm=\lim_{\nu\ra\infty}(2\pi)^J e^{J\nu T/2}\int
\exp\{i\tint_0^T[{\o p}_jd{\o q}^j+d{\o G}({\o p},{\o q})-{\o h}({\o p},
{\o q})dt]\}\,d\mu^\nu_W({\o p},{\o q})\n\\
&&\hskip.3cm=\lim_{\nu\ra\infty}{\cal M_\nu}\int\exp\{i\tint_0^T[{\o p}_j
{\dot{\o q}}^j+{\dot{\o G}}({\o p},{\o q})-{\o h}({\o p},{\o q})dt]\}\n\\
&&\hskip3.2cm\times\exp\{-(1/2\nu)\tint_0^T[d\sigma({\o p},{\o q})^2/dt^2]\,
dt\}\,\D{\o p}\,\D{\o q}\,,
\label{4.4}
\en
where $\o G$ incorporates both $F$ and $G$.
In this expression we have set $d\sigma({\o p},{\o q})^2=dp^2+dq^2$,
namely, the new form of the flat metric in curvilinear phase space
coordinates. We emphasize that this path integral regularization
involves Brownian motion on a flat space whatever
choice of coordinates is made. Our transformation has also made
use of the formal -- and in this case valid -- invariance of the
Liouville measure.

If we have auxiliary terms in the classical action representing
constraints, then the expression of interest would seem to be
\bn
&&\lim_{\nu\ra\infty}{\cal M_\nu}\int\exp\{i\tint_0^T[p_j
{\dot q}^j-h(p,q)-\omega^a\varphi_a(p,q)]\,dt\}\n\\
&&\hskip1cm\times\exp\{-(1/2\nu)\tint_0^T[{\dot p}^2+{\dot q}^2]\,
dt\}\,\D p\,\D q\,\D C(\omega)\;,
\label{4.5}
\en
where the formal measure $\D C(\omega)=\Prod_t\delta\omega(t)$
may be proposed.
We expect some expression of this sort to hold; however,
the explicit proposal in (\ref{4.5}) is incorrect
 as we now proceed to demonstrate.

According to the discussion of the previous sections it is clear
that the physical propagator may also be given by
\begin{equation}
\lim_{\nu\ra\infty}
{\cal M_\nu}\int\limits_{G}^{}\delta\Omega
\int\exp\{i\tint[p_j
{\dot q}^j-h(p,q)]\,dt\}\exp\{-(1/2\nu)\tint{[\dot p}^2+{\dot q}^2]
\,dt\}\,\D p\D q\ ;
\label{4.6}
\end{equation}
here all the paths satisfy $p(T),q(T)=p'',q''$ and $p(0),q(0)=p'^{\,
\Omega},q'^{\,\Omega}$, where following the notation introduced in
Section 2, we define
\bn
p^\Omega=e^{-\Omega^aad\,\varphi_a}p\;,\hskip1.5cm
q^\Omega=e^{-\Omega^aad\,\varphi_a}q\;.
\label{4.7}
\en
In short, we have used the fact that the unitary operators
representing the finite gauge group transformations satisfy
the condition (\ref{c.4})
mapping any coherent state into another coherent state.

Based on the mapping property (\ref{4.7}), we can give another formulation
to the path integral (\ref{4.6}). With the Wiener measure regularization
present, the path integral for any finite $\nu$ is well defined,
and as such we are free to change variables of integration.
In particular, let us make a canonical change of variables so that
\bn
&&p(t)\ra e^{\tint_t^T ds\omega^a(s)ad\,\varphi_a}p(t)\;,\n\\
&&q(t)\ra e^{\tint_t^T ds\omega^a(s)ad\,\varphi_a}q(t)\;,
\label{4.9}
\en
where $\omega^a$ are functions of time subject only to the
requirement that
\bn
\tint_0^T\omega^a(s)\,ds\equiv\Omega^a\;.
\label{4.10}
\en
Clearly, there are infinitely many functions $\omega^a$ that will
satisfy such a criterion, and in a certain sense we will be led to
average over ``all'' of them. Note what this change of variables
accomplishes. In the new variables, whatever the choice of
$\omega^a$ may be, the final values remain unchanged, $p(T),q(T)=p'',
q''$, while the initial values have become $p(0),q(0)=p',q'$ since
$(p'^{\,\Omega})^{-\Omega}\equiv p'$ and $(q'^{\,\Omega})^{-\Omega}
\equiv q'$. Thus we have transformed all the gauge dependence
from the initial points $p'^{\,\Omega},q'^{\,\Omega}$ and have
distributed it throughout the time interval $T$. This discussion is
reminiscent of that in Sections 2 and 3.

It should be remarked that the condition (\ref{4.10})
 may also be avoided if so desired.
Suppose we drop the condition (\ref{4.10}). Let $\bar{\Omega}^a$
be the value of the integral in the right-hand side of (\ref{4.10}).
Since the integral (\ref{4.6}) involves the average over the gauge
orbit that goes through the initial point $p',q'$, the explicit
dependence of the boundary condition on $\bar{\Omega}^a$ at the initial
time can be removed by an appropriate shift of the average parameters
$\Omega^a$. The initial boundary condition remains intact
$p(0),q(0) = p'^\Omega, q'^\Omega$ in contrast to the case when
the condition (\ref{4.10}) is imposed. Nevertheless, we proceed on the
basis of (\ref{4.10}).

Let us next see
what are the consequences for the path integral of such a change
of integration variables. We first observe that
\bn
&&{\dot p}(t)\ra{\dot p}(t)-\omega^aad\,\varphi_ap(t)={\dot p}
(t)-\omega^a\{\varphi_a,p\}(t)\ ,\n\\
&&{\dot q}(t)\ra{\dot q}(t)-\omega^aad\,\varphi_aq(t)={\dot q}(t)-
\omega^a\{\varphi_a,q\}(t)\;.
\label{4.11}
\en
Such a change leads to a new form for the path integral given by
\bn
&&\lim_{\nu\ra\infty}{\cal M_\nu}
\int\limits_{G}^{}\delta\Omega
\int\exp\{i\tint[p_j({\dot q}^j
-\omega^a\{\varphi_a,q^j\})-h(p,q)]\,dt\}\n\\
&&\hskip-1cm\times\exp\{-(1/2\nu)\tint[({\dot p}-\omega^a\{\varphi_a,
p\})^2+({\dot q}-\omega^a\{\varphi_a,q\})^2]\,dt\}\,\D p\,\D q\;.
\label{4.12}
\en
This relation holds because the formal measure remains invariant
under this canonical transformation of coordinates. We recall
that in this form the fixed end points are $p(T),q(T)=p'',q''$
and $p(0),q(0)=p',q'$.
This equation is true for any choice of $\omega^a$ which fulfills
the required integral condition (\ref{4.10}),
and {\it a fortiori} it is
still true if we average (\ref{4.12}) over ``all'' functions which satisfy
the required integral condition. In so doing let us at the same
time incorporate the integral over $\Omega$ and simply average
over ``all'' functions $\omega^a$ directly without any
condition on the overall integral value. For now let us continue
to treat such an average in a formal manner; we will return to
the question of a proper average at a later stage. Thus we may
replace (\ref{4.12}) by
\bn
&&\lim_{\nu\ra\infty}{\cal M_\nu}\int\exp\{i\tint[p_j({\dot q}^j
-\omega^a\{\varphi_a,q^j\})-h(p,q)]\,dt\}
\label{4.13}\\
&&\hskip-1cm\times\exp\{-(1/2\nu)\tint[({\dot p}
-\omega^a\{\varphi_a,p\})^2+({\dot q}-\omega^a\{\varphi_a,q\})^2]\,dt\}\,
\D p\,\D q\,\D C(\omega)\;, \n
\en
where $C(\omega)$ denotes a measure which averages over all
functions $\omega^a$ as required.
Since the object under discussion is manifestly gauge invariant,
it is noteworthy that we can explicitly display such invariance
under the gauge transformations
\bn
\delta p=\{\varphi_a,p\}\delta\lambda^a\;,\hskip1cm
\delta q=\{\varphi_a,q\}\delta\lambda^a\;,\hskip1cm
\delta\omega^a=
\delta\dot{\lambda}^a-f_{abc}\omega^b\delta\lambda_c\;,
\label{4.14}
\en
for general infinitesimal functions $\delta\lambda^a(t)$
which vanish at the end points, and
for which the indicated path integral is invariant for
all values of $\nu$, hence in the limit. Although the
path integral is invariant under the gauge transformations
indicated, the reader should not jump to the conclusion that
the path integral diverges. In fact, the integral over the
gauge functions $\omega^a$ is an {\it average},
that is, $\tint {\cal D}C(\omega)$ is finite, as we have
stressed, and for a bounded integrand no divergences are possible.

Equation (\ref{4.13}) represents a manifestly gauge invariant
expression that is covariant under a general canonical
change of variables. For the class of constraints under
discussion, we can also present another useful expression.
 Using the identity (\ref{c.2}) leads to the equivalent relation
\bn
&&\lim_{\nu\ra\infty}{\cal M_\nu}\int\exp\{i\tint[p_j
 {\dot q}^j -\omega^a\varphi_a(p,q)-h(p,q)]\,dt\}
 \label{4.15}\\
&&\hskip-1cm\times\exp\{-(1/2\nu)\tint_0^T[({\dot p}-\omega^a
\{\varphi_a,p\})^2+({\dot q}-\omega^a\{\varphi_a,q\})^2]\,dt\}\,\D
p\,\D q\,\D C(\omega)\;, \n
\en
 and once again we recognize the parameters $\{\omega^a\}$ as the
Lagrange mutipliers of the classical theory.

Additionally, we observe that the drift terms in the
Wiener measure cannot be neglected. For the Brownian motion
we have the It\^o rule $dp(t)^2=\nu dt$, and the connected expectation value
$E(p(t)p(s))_{\rm conn}=\nu s(1-t/T)$ for $s<t$. Thus the (Stratonovich)
stochastic integral $(1/\nu)\tint \omega^a\{\varphi_a,p\}\,dp$
and the term
$(1/2\nu)\tint[\omega^a\{\varphi_a,p\}]^2\,dt$ are both of order
unity for all values of $\nu$ since in the general case
$\omega^a\{\varphi_a,p\}\simeq p$. A similar discussion holds
for $q$ as well. It is for this reason that our initial naive
proposal (\ref{4.5}) is not acceptable.

\subsubsection*{Choice of measure for the gauge variables}

Finally we take up the question of the choice of the measure
$C(\omega)$ and its associated integral. Although we have
loosely stated that $\D C(\omega)=\Prod_t\delta\omega(t)$ and that
we should integrate over all functions
$\omega^a$, this is still an imprecise concept. Despite appearances, there
is actually a great deal of choice in this measure.
This freedom arises
because the only real requirement on this measure is that it
simulates a {\it single group invariant integral over the initial
parameters} $p'^{\,\Omega},q'^{\,\Omega}$ as discussed in (\ref{4.6}).
We shall consider two possible choices. The first one will provide
us with a manifestly gauge invariant measure, while the second choice
gives an example of a gauge noninvariant measure which nonetheless leads
to the gauge invariant transition amplitude. The latter amounts to
some specific gauge fixing that is manifestly
{\it free} of any Gribov problem.

To define an appropriate measure
that is invariant under general gauge transformations,
we appeal to the classical theory of Kolmogorov \cite{kol}
on stochastic processes, which will ensure that we obtain
a well-defined probability measure on the gauge path space.
Kolmogorov's theorem asserts that an
underlying probability measure on paths exists
provided the set of multi-time joint probability
densities satisfies certain basic consistency conditions.
To show the needed consistency let us again use $\delta\omega
$ as the normalized Haar measure (\ref{haar})
for the compact semi-simple gauge group under consideration.
Then let us introduce a stochastic process defined by the following
set of multi-time joint probability densities
\bn
{\cal P}_n(\omega_n,t_n;\ldots;\omega_2,t_2;\omega_1,t_1)\equiv 1
\label{k.1}
\en
for all $n\geq 1$. Here $T\geq t_n>\cdots t_2>t_1\geq 0$.
The left-hand side of this equation is
the joint probability density for the gauge field
to have value $\omega_1$ at time $t_1$, value $\omega_2$
at time $t_2$, etc.
In this terminology $\omega=\{\omega^a\}$.
The right-hand side of this joint probability density
relation is simply unity, meaning that {\it any} set of values
at {\it any} set of distinct times is equally likely.
This is the proper mathematical statement of a uniform
average over all gauge paths.
Consistency of the given joint probability
densities is simply the trivial
observation that
\bn
&&\int {\cal P}_n(\omega_n,t_n;\ldots;\omega_r,t_r;\dots;\omega_1,t_1)\,
\delta\omega_r\n\\
&&\hskip1cm=1\n\\
&&\hskip1cm={\cal P}_{n-1}(\omega_n,t_n;\ldots;
\omega_{r+1},t_{r+1};\omega_{r-1},t_{r-1};\dots;\omega_1,t_1)\;,
\label{k.2}
\en
for any choice or $r$, $n\geq r\geq 1$, and all $n$,
$n\geq2$; for $n=1$ the last line should be ignored.
The evident consistency of this set of joint
probability densities is then sufficient to guarantee for
us a (countably additive) probability measure
on gauge fields, which we denote by $\rho(\omega)$,
that exhibits these joint probability distributions.

Accepting this choice for the integration over gauge fields
leads to the fact that the physical propagator
may be given the mathematically well-defined formulation
\bn
\hskip-1cm&&\<p'',q''|e^{-i\H T}|p',q'\>^{ph}\n\\
&&=\lim_{\nu\ra\infty}(2\pi)^J
e^{J\nu T/2}\int\exp\{i\tint[p_jdq^j+dG(p,q)
-\omega^a\varphi_a(p,q)dt-h(p,q)dt]\}\n\\
&&\hskip3.5cm\times\,d\mu^\nu_W(p,q,\omega)\,d\rho
(\omega)\;;
\label{k.3}
\en
here we have added $\omega$ to the argument of $\mu^\nu_W$
to acknowledge the presence of the drift terms.
The result only depends on the initial and final values of
$p$ and $q$ since we have integrated over the set of gauge paths
without any boundary conditions;
this result is still invariant under
continuous and differential gauge transformations (\ref{4.14}).

Finally we note that the relation between
the physical Hamiltonian operator and
the classical expression $h(p,q)$ is given by
\bn
{\cal H}_{ph}\equiv
\int h(p,q)\,|p,q\>^{ph}\,^{ph}
\<p,q|\,d^J\!p\,d^J\!q/(2\pi)^J\;,
\label{k.4}
\en
where the physical coherent state $|p,q\>^{ph}$ is obtained
by the average of the coherent state (\ref{c.4}) over the group
$G$ with the normalized measure $\delta\Omega$.

Formally, the measure $d\rho(\omega)$ constructed above comes naturally
from the convolution formula (\ref{3.1}) where each infinitesimal
transition amplitude is to be replaced by the corresponding infinitesimal
amplitude (\ref{4.6}) with the Wiener measure. In this construction
the projection operator is inserted at each moment of time, that is, formally,
$d\rho(\omega) = \prod_t \delta \omega(t)$. Clearly, this formal measure
satisfies the conditions (\ref{k.1}) and (\ref{k.2}), and in addition it is
manifestly gauge invariant and normalized $\tint d\rho(\omega) =1$.

  However, from the calculational point of view the measure $\rho(\omega)$
is not always convenient. Sometimes it is also useful to have
a measure for the gauge variables that is not explicitly gauge
invariant (gauge fixing). A conventional gauge fixing discussed
in Section 3 may suffer from Gribov ambiguities. Next we
show an example of a Gaussian probability measure free of such a disease.

Since we want the measure to have at least one average over
the group manifold $G$, it is natural to assume
that for any time slice the measure must be the  group
invariant measure, but what is at our disposal is the
relationship of the functions at neighboring points of time.
As one set of examples, it would suffice to restrict our
integration to the set, or even a subset, of {\it continuous
functions}. A natural way to achieve it is to choose $\D C(\omega)$
to be a Wiener measure on the manifold $G$
\bn
\D C(\omega)= d\rho_W(\omega)=
{\cal N}\exp[-\half\tint g_{ab}(\omega)
{\dot\omega}^a{\dot\omega}^b\,dt]\,\Pi_t\,\delta\omega(t)\;.
\label{b.1}
\en
Here the metric $g_{ab}(\omega)$ is the positive-definite
metric associated with a homogeneous space determined by the
compact semi-simple gauge group.
The measure can also be regarded as the imaginary time quantum dynamics
 of a free particle propagating on the compact
homogeneous manifold $G$.

Let us now establish a relation between the projection formula
(\ref{4.6}) and (\ref{4.15}) with the choice (\ref{b.1})
for the measure. Let $g_\omega$ be an element of the gauge group in a
matrix representation. Then the action in the exponential
in (\ref{b.1}) can also be rewritten as
\begin{equation}
S_W =-c\,{\rm tr}\int\limits_{0}^{T}(\dot{g}_\omega
g_\omega^{-1})^2/2dt\ ,
\label{b.2}
\end{equation}
where $c=1/{\rm tr}(1)$ is a normalization factor.
Consider a transition amplitude of
a free particle on the manifold $G$
\begin{equation}
K_T(g_\Omega, g_{\Omega'}) ={\cal N}
\int\limits_{g_\omega(0)=g_{\Omega'}}^{g_\omega(T)= g_\Omega}
\prod\limits_{t=0}^T \delta\omega(t) e^{-S_W}\ ,
\label{b.3}
\end{equation}
normalized so as to satisfy
\begin{equation}
K_T(g_{\Omega''},g_{\Omega'})=\int K_{T-t}
 (g_{\Omega''},g_\Omega)K_t(g_\Omega,g_{\Omega'})\,\delta\Omega\,.
\label{b.3a}
\end{equation}
Due to the global invariance of the action with respect to the left and
right shifts, $g_\omega\rightarrow g_0g_\omega$ and
$g_\omega\rightarrow g_\omega g_0$,  the amplitude (\ref{b.3}) is
also invariant under these transformations
\begin{equation}
K_T(g_\Omega, g_{\Omega'}) =
K_T(g_0g_\Omega,g_0 g_{\Omega'}) =
K_T(g_\Omega g_0, g_{\Omega'}g_0) \ .
\label{b.4}
\end{equation}
From (\ref{b.4}) we deduce the identity
\begin{equation}
\int\limits_{G}^{}\delta\Omega''
K_T(g_{\Omega''}, g_{\Omega'}) =
\int\limits_{G}^{}\delta\Omega'
K_T(g_{\Omega''}, g_{\Omega'}) =  1\ ,
\label{b.5}
\end{equation}
 which can be easily seen from the Feynman-Kac representation
of the transition amplitude (\ref{b.3}) as a spectral sum.
The integral (\ref{b.5}) determines an action of the evolution
operator on the ground state of the system. So, only the ground
state will contribute to the integral. We naturally assume that
the Casimir energy (the ground state energy) can always be
subtracted and included into the path integral normalization.

Now we insert the identity (\ref{b.5}) into the measure of the
path integral (\ref{4.6}) and then proceed with the change
of variables (\ref{4.9}). Since in the identity (\ref{b.5})
either $\Omega''$ or $\Omega'$ is a free parameter, we can
always choose it to coincide with the parameter $\Omega$
of the $G$-average in (\ref{4.6}). Substituting
the path integral representation of $K_T$ (\ref{b.3})
in the appropriately transformed integral (\ref{4.6}), we arrive
at the expression (\ref{4.15}) with the Wiener measure
(\ref{b.1}) for the gauge variables.

Typically we
encounter Wiener measures that are pinned at either the
initial time or at both end points; in the present case,
the measure for gauge variables
is neither pinned at the initial nor the final
time as seen from the derivation of (\ref{4.15}).
Since the group is compact, the group volume is finite
and we may therefore normalize such a Wiener measure that is
not pinned; our normalization is such that
\begin{equation}
\int\D C(\omega)=
\int\limits_{G}^{}\delta\Omega''\delta\Omega'
K_T(g_{\Omega''}, g_{\Omega'}) =  1\ .
\label{b.6}
\end{equation}
In that case the formal measure $\D C(\omega)$ is actually a
well-defined (countably additive) probability measure which
we denote by $d\rho_W(\omega)$.
With this choice we note that
the physical propagator may also be given the well-defined
definition (\ref{k.3}) where $d\rho(\omega)\rightarrow
d\rho_W(\omega)$.
The result only depends on the
initial and final values of $p,q$ since we have integrated
over the set of continuous $\omega^a$ paths without any
boundary conditions.

The measure is not invariant under the gauge transformations
(\ref{4.14}), nonetheless the transition amplitude is gauge
invariant because the measure provides the necessary projection
onto gauge invariant states. In contrast to the conventional
procedure of section 3, there is no explicit gauge condition
imposed on the system of phase space variables, and hence the
Gribov problem is avoided.

One should add that two such propagators, one from $t=0$ to
$t=T$ and the second from $t=T$ to $t=2T$, for example,
seems to
not compose to a propagator of the same form as (\ref{k.3})
due to the discontinuity of paths at the interface.
However, the resultant propagator  is nonetheless correct;
it simply involves another acceptable form for the measure
${\cal D}C(\omega)$.

\subsubsection*{Conclusion}

With (\ref{4.15}) and two  choices
of the measure for the gauge variables, we have arrived at
our coordinate-free and mathematically
well-defined formulation for the path integral representation of
the special class of first-class constraints that was our goal.

\end{document}